%% file: article.tex
\documentclass[article,12pt]{elsarticle}

\usepackage[utf8]{inputenc} 
\usepackage{tablefootnote}
\usepackage{longtable}
\usepackage{url}

\usepackage{pgfplots}
\pgfplotsset{width=7cm,compat=1.8}
\usepackage{pgfplotstable}

\usepackage{sfmath}
\usepackage{amssymb}

\begin{document}

\include{content}

\end{document}

%% file: content.tex
\section*{Coverage of Author Identifiers in Web of Science and Scopus}

Thomas Krämer\footnote{Corresponding author: thomas.kraemer@gesis.org}, Fakhri Momeni, Philipp Mayr

\textit{GESIS -- Leibniz-Institute for the Social Sciences, Cologne, Germany}

\section*{Abstract}
As digital collections of scientific literature are widespread and used frequently in knowledge-intense working environments, it has become a challenge to identify author names correctly. 
The treatment of homonyms is crucial for the reliable resolution of author names. Apart from varying handling of first, middle and last names, vendors as well as the digital library community created tools to address the problem of author name disambiguation.
This technical report focuses on two widespread collections of scientific literature, Web of Science (WoS) and Scopus, and the coverage with author identification information such as Researcher ID, ORCID and Scopus Author Identifier in the period 1996 - 2014.
The goal of this study is to describe the significant differences of the two collections with respect to overall distribution of author identifiers and its use across different subject domains. We found that the STM disciplines show the best coverage of author identifiers in our dataset of 6,032,000 publications which are both covered by WoS and Scopus. In our dataset we found 184,823 distinct ResearcherIDs and 70,043 distinct ORCIDs. In the appendix of this report we list a complete overview of all WoS subject areas and the amount of author identifiers in these subject areas.

\textbf{Keywords}: Disambiguation; author name homonyms; author identification; citation databases; WoS subject areas; Researcher ID; ORCID; Scopus Author Identifier

\section{Introduction}
\label{Introduction}

The role of unique identifiers in the scientific publishing infrastructure is becoming more and more important \cite{Kraker2015}. Especially author identifiers \cite{Warner2010} play a crucial role in bibliometric research and other searching and harvesting activities.

Author name disambiguation has an important effect on bibliometric analysis \cite{Reijnhoudt2013,Ioannidis2014}, scholarly networks \cite{diesneretal} and all other scientometric approches using large-scale bibliographic data \cite{Mutschke2015}. Therefore an author identification system with a high accuracy is vital for the further processing on bibliometric data. Harzing \cite{HealthWarning} explains how frequency of homonym names in Thomson Reuter’s Essential Science Indicators can be problematic. Also, Torvik and Smalheiser \cite{Torvik} mentioned that accuracy of author disambiguation depends on name frequency and common names tend to be harder to disambiguate. Kawashima and Tomizawa \cite{Kawashima} evaluated the accuracy of Scopus author identifier based on the largest funding database in Japan. They got a high precision and recall, but they noted the dependency of author ID accuracy on country or language. 

Several approaches exist to address the systemic name ambiguity problems in research publications:
automated or analytical procedures that apply algorithms to publication metadata to identify and group publications of the same author as well as manual approaches that require author interaction to claim or verify authorship \cite{DeCock2013}. Three recent surveys about author name disambiguation have been presented in the literature \cite{Ferreira2012,Elliott2010,Smalheiser2009}. In these surveys the basic problems of author name disambiguation are outlined and the most popular methods are overviewed.

Researchers can register their ResearcherID by self-registration either as subscribers of Web of Science or at the ResearcherID website\footnote{http://www.researcherid.com/}. The link between documents in Web of Science and ResearcherIDs is established by means of a publication list: Authors themselves add publications to their profile and thus connect their ResearcherID with the publication metadata in Web of Science. 

In Scopus, Scopus Author Identifiers are assigned automatically. The algorithm groups author names under a common Scopus Author Identifier based on an algorithm that matches affiliation, address, subject area, source title, dates of publication citations, and co-authors information \cite{scopusauthorid}. However, the exact matching procedure of Scopus is not known. 
ResearcherID and Scopus Author Identifier are integrated into their respective digital libraries, Web of Science resp. Scopus. 

Independent from these two systems, ORCID\footnote{http://orcid.org/} is an open system designed for author identification \cite{Haak2012}. The system started in 2012 and is operated by a non-for-profit organization whose members are commercial as well as educational institutions. Similar to ResearcherID, authors register themselves and receive an identifier which then should be used whenever work is submitted for publication. A couple of journals (e.g. published with Springer or PLOS) use the ORCID identifier already in the submission process of a manuscript and publish the ORCID together with the other author information. 

In this technical report we describe the existence of three prominent author identifiers (ResearcherID, ORCID and Scopus Author Identifier) \cite{fenner2011} in a joint subset of the Web of Science and Scopus. 
We will report about the gaps in coverage in different WoS subject areas. 
Finally we will try to show the effect of author name homonyms on the distribution of ORCIDs (orcid), ResearcherIDs (rid) and Scopus Author identifiers (Scopus).

\section{Dataset}\label{sec:dataset}
The Competence Centre for Bibliometrics\footnote{http://www.bibliometrie.info/} in Germany has access to the raw data exports of Scopus and Web of Science from January 1996 to April 2014. Based on the raw exports, Leibniz Institute for Information Infrastructure (FIZ Karlsruhe)\footnote{https://www.fiz-karlsruhe.de/en.html} has created an infrastructure for bibliometric analyses that contains standardized representations of both collections.
This infrastructure provides a duplicate database\footnote{Here, duplicate denotes publications that are present in both WoS and Scopus database while \cite{RePEc:eee:infome:v:9:y:2015:i:3:p:570-576} conducted duplicate record detection within Scopus metadata.} that maps pairs of identical publications. 
FIZ Karlsruhe automatically identified duplicates by pairwise comparisons of WoS and Scopus metadata. Each pair was checked multiple times, applying different properties to be compared (see Table \ref{fiz:criteria}). 
We decided to use a rather strict criterion to define the common subset for our analysis in order to achieve two standardized, nearly identical samples of publications in both collections. We set a minimum threshold of 317 (compare Table \ref{fiz:criteria}) to fetch documents from both collections.
We consider pairs of metadata from WoS and Scopus describing an identical publication only if the DOI and two or more other metadata fields match in both records. Other fields are publisher item identifier, issue, publication year, volume, ISSN, first page, first author or title.
Applying these criteria we got a dataset that consists of pairs of metadata describing approximately 6,032,000 publications which are both covered by WoS and Scopus.

\begin{table}[!htb]
	\caption{Weights assigned to pairs of metadata according to matching properties}\label{fiz:criteria}
	\begin{center}	%
		\begin{tabular}{r|l}
			  weight & criteria                                               \\ \hline
			$\vdots$ & $\vdots$                                               \\ \hline
			      93 & Title, 1.Page, ISSN, Volume, Issue                     \\ \hline
			$\vdots$ & $\vdots$                                               \\ \hline
			     317 & DOI, 1.Author, 1.Page, ISSN, Volume, Issue             \\ \hline
			$\vdots$ & $\vdots$                                               \\ \hline
			     509 & DOI, PII, Title, 1.Author, 1.Page, ISSN, Volume, Issue \\ \hline
		\end{tabular} 
	\end{center}
\end{table}

A small difference (below 0.02\%) in the number of publications indicates that the chosen criteria are appropriate to identify identical publications in the two collections.

\section{Method}\label{sec:method}

\subsection{Comparing subject areas}	

We chose the WoS subject areas\footnote{http://incites.isiknowledge.com/common/help/h\_field\_category\_wos.html} as reference for the following comparison. As the duplicate database provides a reliable means to identify one specific document in both corpora, it is feasible to apply the subject areas of WoS to the same document in the Scopus database.  
We grouped all documents in our subset according to the publication year and the WoS subject areas that have been assigned to the same document.

\subsection{Checking author homonyms}
As another selection criteria we chose those first and last name pairs that have either an author id, an orcid or a researcher id assigned. 
We assumed that a large number of institutions and a very wide range of journals per identifiable author could be the result of incorrect author id assignment. 
For each first and last name pair with a single identifier of one of the three author id systems and a minimum number of 100 publications, we checked the distributions of institutions they have been affiliated to and the number of journals in which the related publication appeared.

\section{Results}\label{sec:result}

\subsection{General coverage of author identifiers}
Our analysis leads to 184,823 distinct ResearcherIDs (rid) and 70,043 distinct ORCIDs. In the Scopus data, we found 9,839,749 authors, out of which 9,839,711 (100 \%) have a Scopus Author ID (authorid) assigned. The automatic assignment of Scopus Author Identifiers succeeded in nearly every document (see Table \ref{tab:sco-aid}) in the subset, whereas the relative number of ResearcherIDs related to publications in recent years decreased steadily (see Table \ref{tab:april}). 

\begin{table}[!htb]
	\caption{ORCID and ResearcherIDs - yearly percentage \tablefootnote{The publication year 2014 includes data from January to April.}\label{tab:april}}
	\begin{center}
		\begin{tabular}{r|r|r|r|r|rl}
			year & authors in documents &     rid & \%  rid &   orcid & \%  orcid &  \\ \hline
			2014 &            1,267,886 &  30,706 &    2.42 &  17,681 &      1.39 &  \\ \hline
			2013 &            6,207,543 & 213,368 &    3.44 & 111,535 &      1.80 &  \\ \hline
			2012 &            5,706,042 & 274,630 &    4.81 & 123,964 &      2.17 &  \\ \hline
			2011 &            5,083,237 & 294,323 &    5.79 & 119,752 &      2.36 &  \\ \hline
			2010 &            4,508,560 & 281,554 &    6.24 & 108,953 &      2.42 &  \\ \hline
			2009 &            4,052,351 & 278,685 &    6.88 & 105,504 &      2.60 &  \\ \hline
			2008 &            2,970,989 & 218,877 &    7.37 &  81,940 &      2.76 &  \\ \hline
			2007 &              250,397 &  18,946 &    7.57 &   7,305 &      2.92 &  \\ \hline
			2006 &               13,791 &     773 &    5.61 &     309 &      2.24 &  \\ \hline
			2005 &                6,640 &     393 &    5.92 &     157 &      2.36 &  \\ \hline
			2004 &                4,728 &     332 &    7.02 &     162 &      3.43 &  \\ \hline
			2003 &                2,198 &     120 &    5.46 &      57 &      2.59 &  \\ \hline
			2002 &                1,429 &      90 &    6.30 &      45 &      3.15 &  \\ \hline
			2001 &                1,162 &      99 &    8.52 &      42 &      3.61 &  \\ \hline
			2000 &                  451 &      19 &    4.21 &      11 &      2.44 &  \\ \hline
			1999 &                  102 &       4 &    3.92 &       2 &      1.96 &  \\ \hline
			1998 &                   58 &       2 &    3.45 &       1 &      1.72 &  \\ \hline
			1997 &                   90 &      10 &   11.11 &       4 &      4.44 &  \\ \hline
			1996 &                  165 &       8 &    4.85 &       3 &      1.82 &  \\ \hline
		\end{tabular}
	\end{center}
\end{table}

\begin{table}[!h]
	\caption{Scopus author ids - yearly percentage}  \label{tab:sco-aid} 
	\begin{center}
		\begin{tabular}{r|r|r|r|r}
			year &  authorid &     total &  \% & missing values \\ \hline
			2014 & 1,262,894 & 1,262,894 & 100 &              0 \\ \hline
			2013 & 6,312,343 & 6,312,343 & 100 &              0 \\ \hline
			2012 & 5,929,758 & 5,929,758 & 100 &              0 \\ \hline
			2011 & 5,275,211 & 5,275,214 & 100 &              3 \\ \hline
			2010 & 4,704,713 & 4,704,715 & 100 &              2 \\ \hline
			2009 & 4,258,223 & 4,258,250 & 100 &             27 \\ \hline
			2008 & 3,093,222 & 3,093,228 & 100 &              6 \\ \hline
			2007 &   258,001 &   258,001 & 100 &              0 \\ \hline
			2006 &    14,673 &    14,673 & 100 &              0 \\ \hline
			2005 &     6,933 &     6,933 & 100 &              0 \\ \hline
			2004 &     4,865 &     4,865 & 100 &              0 \\ \hline
			2003 &     2,943 &     2,943 & 100 &              0 \\ \hline
			2002 &     1,494 &     1,494 & 100 &              0 \\ \hline
			2001 &     1,262 &     1,262 & 100 &              0 \\ \hline
			2000 &       466 &       466 & 100 &              0 \\ \hline
			1999 &       109 &       109 & 100 &              0 \\ \hline
			1998 &       108 &       108 & 100 &              0 \\ \hline
			1997 &        91 &        91 & 100 &              0 \\ \hline
			1996 &       169 &       169 & 100 &              0 \\ \hline
		\end{tabular}	
	\end{center}				
\end{table}
\clearpage

\subsection{Coverage in subject areas}
We summed the absolute numbers of author ids (any system) present in the subset. Within the twenty most frequent subject areas, Scopus author identifiers are particularly dominant in medical domains such as 'oncology', 'surgery' or 'immunology' (see Figure 1).

\includegraphics[width=13cm]{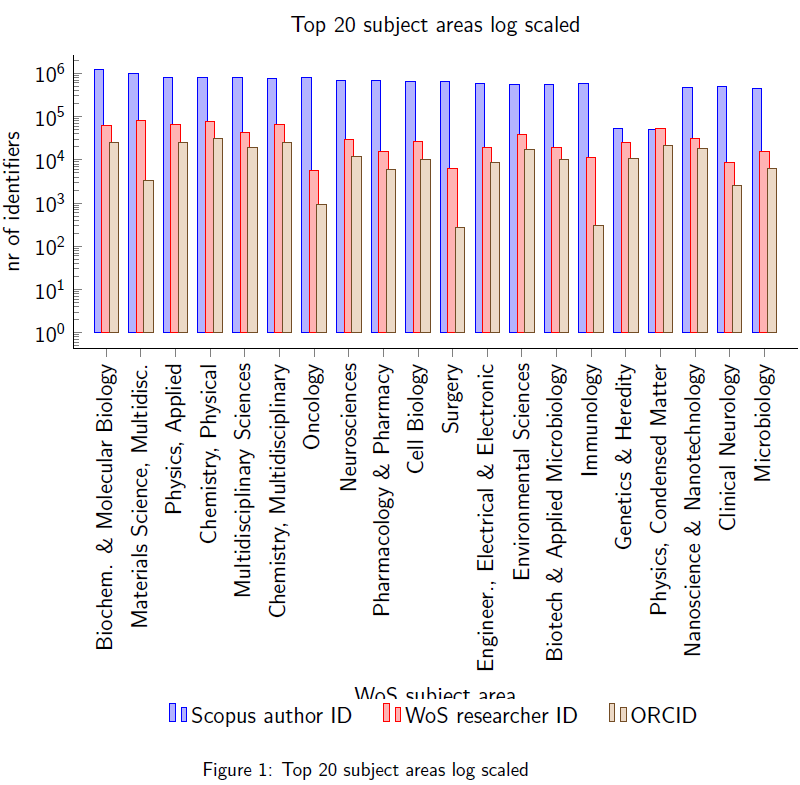}

\begin{table}[!h]
	\caption{Top 10 subject areas covered with researcher ids}
	\begin{center}	
		\begin{tabular}{r|r}
			                      classification & researcher id \\ \hline
			Materials Science, Multidisciplinary &         65,837 \\ \hline
			                 Chemistry, Physical &         64,639 \\ \hline
			                    Physics, Applied &         61,655 \\ \hline
			        Chemistry, Multidisciplinary &         52,875 \\ \hline
			Biochemistry \&    Molecular Biology &         46,393 \\ \hline
			           Physics, Condensed Matter &         42,387 \\ \hline
			 Nanoscience \&       Nanotechnology &         39,123 \\ \hline
			          Multidisciplinary Sciences &         30,250 \\ \hline
			              Environmental Sciences &         29,394 \\ \hline
			          Physics, Multidisciplinary &         29,141 \\ \hline
		\end{tabular}
	\end{center}
\end{table}
\begin{table}[!h]
	\caption{Top 10 subject areas covered with orcids}
	\begin{center}
		\begin{tabular}{r|r}
			                      classification & orcid \\ \hline
			Materials Science, Multidisciplinary & 32,940 \\ \hline
			                 Chemistry, Physical & 30,708 \\ \hline
			                    Physics, Applied & 25,854 \\ \hline
			        Chemistry, Multidisciplinary & 25,382 \\ \hline
			   Biochemistry \& Molecular Biology & 24,721 \\ \hline
			           Physics, Condensed Matter & 21,306 \\ \hline
			          Multidisciplinary Sciences & 19,002 \\ \hline
			       Nanoscience \& Nanotechnology & 18,271 \\ \hline
			              Environmental Sciences & 17,694 \\ \hline
			          Physics, Multidisciplinary & 13,166 \\ \hline
		\end{tabular}
	\end{center}
\end{table}

\clearpage
\begin{table}[!t]
	\caption{Top 10 subject areas covered with Scopus author ids}
	\begin{center}
		\begin{tabular}{r|r}
			                       classification & Scopus author id \\ \hline
			    Biochemistry \& Molecular Biology &         1,256,720 \\ \hline	
			 Materials Science, Multidisciplinary &           983,555 \\ \hline
			                     Physics, Applied &           822,106 \\ \hline
			           Multidisciplinary Sciences &           820,757 \\ \hline
			                  Chemistry, Physical &           798,542 \\ \hline
			                             Oncology &           792,495 \\ \hline
			         Chemistry, Multidisciplinary &           747,621 \\ \hline
			                        Neurosciences &           699,094 \\ \hline
			Pharmacology \&              Pharmacy &           685,833 \\ \hline
			                         Cell Biology &           664,492 \\ \hline
		\end{tabular}
	\end{center}
\end{table}

\subsection{Identification of homonyms}
The following tables show the total, average and maximum number for each system with author ids related to at least 100 publications in the subset\footnote{Please note that the descriptive statistics for publication (see Table \ref{hmnm:total}),  for institution (Table \ref{hmnm:inst}) and issue distribution (see Table \ref{hmnm:journal}) rely on the primary keys assigned to an institution resp. issue in the FIZ Karlsruhe bibliographic database.}. 

\begin{table}[!h]			
	\caption{Publications of supposed frequent authors}\label{hmnm:total}
	\begin{center}
		\begin{tabular}{r|r|r|r}
			& orcid &   rid & Scopus \\ \hline
			total author ids &    91 &    86 &  8,338 \\ \hline
			average publications per author & 132.7 & 133.6 &  142.2 \\ \hline
			publications maximum &   355 &   355 &  2,338
		\end{tabular} 
	\end{center}
\end{table}	

We found 91 authors with an ORCID, that have 100 or more publications, with 132.7 as average and 355 as the maximum. 
The Scopus data contains 8,338 authors that are each related to 100 or more publications, and one Scopus Author Identifier appearing in 2,338 publications (see Table \ref{hmnm:total}).

\pagebreak
The number of publications and journals of these supposed frequent writers are similar for related ResearcherIDs and ORCIDs.

\begin{table}[!h]			
	\caption{Journals in which supposed frequent writers appear}\label{hmnm:journal}
	\begin{center}
		\begin{tabular}{r|r|r|r}
			                             & orcid &   rid & Scopus \\ \hline
			                  author ids &    90 &    81 &  8,219 \\ \hline
			appeared in journals average & 115.1 & 115.5 &  118.4 \\ \hline
			appeared in journals maximum &   329 &   329 &  2,322
		\end{tabular} 
	\end{center}
\end{table}		   

This does not apply to the number of institutions associated with these publications, which is less for ResearcherID. 

For Scopus, all distributions have much higher maximum values, while the averages differ significantly for the institutions related to the supposed frequent writers: 8,219 Scopus Identifiers have 100 or more publications, that are related to 151.4 institutions on average (see Table \ref{hmnm:inst}).

\begin{table}[!h]			
	\caption{Institutions supposed frequent writers are associated with}\label{hmnm:inst}
	\begin{center}
		\begin{tabular}{r|r|r|r}
			                                     & orcid &  rid & Scopus \\ \hline
			                          author ids &    90 &   81 &  8,219 \\ \hline
			affiliated with institutions average &  98.6 & 87.0 &  151.4 \\ \hline
			affiliated with institutions maximum & 1,417 &  344 &  5,364
		\end{tabular} 
	\end{center} 
\end{table}

\pagebreak
For each of the three author identification systems, we chose ten authors that were related to the maximum number of publications. Tables \ref{homonyms:orcid} and \ref{homonyms:scopus} show a limitation of the automated procedure applied in Scopus:
While the supposed top ten list of author names for ResearcherID and ORCID do contain Western as well as Asian names, the top ten Scopus Author Identifiers contain exclusively Asian names, which have a common Western transliteration. 

\begin{table}[!h]			
	\caption{Names of 10 authors with highest publication count - orcid}
	\label{homonyms:orcid}
	\begin{center}		
		\begin{tabular}{r|r|r|r}
			      first &          last &               orcid & publications \\ \hline
			    PAUL K. &           CHU & 0000-0002-5581-4883 &          355 \\ \hline
			  ULRICH S. &      SCHUBERT & 0000-0003-4978-4670 &          270 \\ \hline
			ABDULLAH M. &         ASIRI & 0000-0001-7905-3209 &          210 \\ \hline
			  KRZYSZTOF & MATYJASZEWSKI & 0000-0003-1960-3402 &          193 \\ \hline
			 RICHARD D. &         SMITH & 0000-0002-2381-2349 &          183 \\ \hline
			      HIDEO &        HOSONO & 0000-0001-9260-6728 &          179 \\ \hline
			    AKIHISA &         INOUE & 0000-0002-7546-5334 &          167 \\ \hline
			      JINDE &           CAO & 0000-0003-3133-7119 &          163 \\ \hline
			         L. &    SCODELLARO & 0000-0002-4974-8330 &          163 \\ \hline
			    ANNE B. &        NEWMAN & 0000-0002-0106-1150 &          160
		\end{tabular}				
	\end{center}				
\end{table}

\begin{table}[!h]			
	\caption{Names of 10 authors with highest publication count - Scopus author identifier}\label{homonyms:scopus}
	
	\begin{center}				
		\begin{tabular}{r|r|r|r}
			first & last &   author id & publications \\ \hline
			  WEI & WANG &  7501761836 &        2,338 \\ \hline
			  WEI &  LIU & 36077269600 &        1,773 \\ \hline
			  YAN & WANG &  8640586600 &        1,735 \\ \hline
			 YING & WANG & 55757783062 &        1,448 \\ \hline
			  WEI & CHEN &  9639099600 &        1,375 \\ \hline
			   LI &   LI & 55218828300 &        1,285 \\ \hline
			  WEI & WANG & 36072896000 &        1,218 \\ \hline
			  HUI &   LI & 56002811300 &        1,198 \\ \hline
			QIANG & WANG & 35178491000 &        1,146 \\ \hline
			   XI & CHEN &  8043429800 &        1,105
		\end{tabular}				
	\end{center}				
\end{table}

\pagebreak
\section{Discussion and conclusions }\label{sec:conclusion}
In this report we compared Web of Science and Scopus with regard to different author identifier systems. 
Documents in WoS and Scopus were considered identical if DOI and at least two more criteria matched. 
This might introduce artifacts, as some disciplines might have adopted DOI assignment earlier than others and are therefore overrepresented.

ORCID and ResearcherID require human intervention while Scopus Author Identifier is assigned automatically. 
Comparing documents with a manually assigned identifier system, ResearcherID is assigned more often (70.5\%) than ORCID (29.5\%).

Asian name transliteration leads to common representations of originally distinct author names \cite{Wang745}. 
This, in turn, results in a remarkable loss of precision of many automated author matching procedures \cite{Strotmann2012}. 
Therefore, Scopus Author Identifier is not a reliable source for author assignments. Elsevier itself promotes the use of ORCID instead and offers the possibility to import publication metadata from Scopus to ORCID\footnote{https://orcid.scopusfeedback.com/}. Authors might sort out false positives during this step and help to improve the Scopus Author Identifier assignment. 

The review of author identifiers and the assignment of classification terms points into the same direction. The manual approach ThomsonReuters chose with ResearcherID leads to less ambiguous and more reliable author information, but it requires engagement of the authors. 
In order to minimize the effort needed to integrate with ORCID, users can link the ResearcherID platform with their ORCID account\footnote{http://wokinfo.com/researcherid/integration} to send their publication list from ResearcherID to ORCID and vice versa. 
ORCID is focussed solely on providing persistent identifiers for researchers.
The ORCID technology is open source and the web platform can be integrated into other services using various APIs\footnote{https://orcid.org/organizations/integrators/API}. This openness might be crucial to encourage researchers and service providers to adopt ORCID. 

We showed, that at the time of writing, the majority of researchers do not use ORCID or ResearcherID. The steadily growing number of ORCID users indicates that this might change in the future, but as long as this is the case it is necessary to elaborate automated author disambiguation technologies for resilient bibliometrics and increase their accuracy \cite{Momeni2016,PMID:doiInHumSoc}.

\section{Acknowlegdement}

This work was partly funded by BMBF (Federal Ministry of Education and Research, Germany) under grant number 01PQ13001. We thank your colleagues at the Competence Centre for Bibliometrics and Leibniz Institute for Information Infrastructure (FIZ Karlsruhe) for providing access and assistance.



\bibliographystyle{elsarticle-harv}
\bibliography{authoridsystems}

\clearpage
\section{Appendix}

\subsection{Author identifiers in WoS subject areas }

\begin{longtable}{|l|r|r|r|} 
	\hline
	Subject area & Scopus & rid & orcid \\
	\hline
	\endhead
	Biochemistry \& Molecular Biology                & 1,256,720 & 61,655 & 24,721 \\ \hline
	Materials Science, Multidisciplinary             &  983,555  & 82,838 & 32,940 \\ \hline
	Physics, Applied                                 &  822,106  & 65,837 & 25,854 \\ \hline
	Chemistry, Physical                              &  798,542  & 76,455 & 30,708 \\ \hline
	Multidisciplinary Sciences                       &  820,757  & 42,387 & 19,002 \\ \hline
	Chemistry, Multidisciplinary                     &  747,621  & 64,639 & 25,382 \\ \hline
	Oncology                                         &  792,495  & 17,247 & 6,174  \\ \hline
	Neurosciences                                    &  699,094  & 29,141 & 12,007 \\ \hline
	Pharmacology \& Pharmacy                         &  685,833  & 22,422 & 8,983  \\ \hline
	Cell Biology                                     &  664,492  & 27,041 & 10,428 \\ \hline
	Surgery                                          &  652,405  & 9,574  & 3,340  \\ \hline
	Engineering, Electrical \& Electronic            &  586,457  & 29,394 & 12,475 \\ \hline
	Environmental Sciences                           &  548,327  & 39,123 & 17,694 \\ \hline
	Biotechnology \& Applied Microbiology            &  555,379  & 27,943 & 11,852 \\ \hline
	Immunology                                       &  569,124  & 16,026 & 5,939  \\ \hline
	Genetics \& Heredity                             &  541,310  & 25,625 & 10,575 \\ \hline
	Physics, Condensed Matter                        &  498,070  & 52,875 & 21,306 \\ \hline
	Nanoscience \& Nanotechnology                    &  467,866  & 46,393 & 18,271 \\ \hline
	Clinical Neurology                               &  499,098  & 13,918 & 5,481  \\ \hline
	Microbiology                                     &  444,834  & 18,981 & 7,808  \\ \hline
	Physics, Multidisciplinary                       &  378,029  & 30,250 & 13,166 \\ \hline
	Medicine, Research \& Experimental               &  403,176  & 11,204 & 4,405  \\ \hline
	Public, Environmental \& Occupational Health     &  401,694  & 11,766 & 5,040  \\ \hline
	Endocrinology \& Metabolism                      &  400,239  & 12,950 & 5,134  \\ \hline
	Chemistry, Organic                               &  381,821  & 20,184 & 7,981  \\ \hline
	Radiology, Nuclear Medicine \& Medical Imaging   &  391,466  & 11,826 & 4,814  \\ \hline
	Chemistry, Analytical                            &  366,278  & 22,824 & 9,627  \\ \hline
	Cardiac \& Cardiovascular System                 &  388,487  & 7,291  & 2,753  \\ \hline
	Biochemical Research Methods                     &  358,469  & 21,718 & 9,318  \\ \hline
	Engineering, Chemical                            &  356,641  & 21,956 & 9,406  \\ \hline
	Plant Sciences                                   &  336,511  & 20,052 & 8,535  \\ \hline
	Hematology                                       &  339,350  & 8,611  & 3,164  \\ \hline
	Medicine, General \& Internal                    &  339,698  & 6,950  & 2,721  \\ \hline
	Biophysics                                       &  307,776  & 18,170 & 7,460  \\ \hline
	Infectious Diseases                              &  312,904  & 7,732  & 3,002  \\ \hline
	Food Science \& Technology                       &  302,714  & 13,962 & 6,477  \\ \hline
	Pediatrics                                       &  313,091  & 5,361  & 1,931  \\ \hline
	Energy \& Fuels                                  &  290,779  & 17,550 & 7,686  \\ \hline
	Chemistry, Medicinal                             &  298,768  & 10,729 & 4,246  \\ \hline
	Biology                                          &  288,089  & 17,706 & 7,538  \\ \hline
	Instruments \& Instrumentation                   &  288,854  & 16,401 & 6,966  \\ \hline
	Geosciences, Multidisciplinary                   &  277,734  & 23,088 & 10,176 \\ \hline
	Optics                                           &  277,831  & 19,587 & 8,219  \\ \hline
	Psychiatry                                       &  287,043  & 10,828 & 4,674  \\ \hline
	Polymer Science                                  &  273,400  & 20,009 & 7,798  \\ \hline
	Physics, Atomic, Molecular \& Chemical           &  251,968  & 29,009 & 12,346 \\ \hline
	Gastroenterology \& Hepatology                   &  284,031  & 5,214  & 1,863  \\ \hline
	Astronomy \& Astrophysics                        &  263,826  & 12,110 & 5,284  \\ \hline
	Ecology                                          &  244,281  & 25,280 & 11,310 \\ \hline
	Peripheral Vascular Diseases                     &  270,018  & 6,621  & 2,456  \\ \hline
	Chemistry, Applied                               &  257,403  & 13,591 & 5,969  \\ \hline
	Physiology                                       &  234,179  & 9,889  & 3,849  \\ \hline
	Toxicology                                       &  233,588  & 9,398  & 3,873  \\ \hline
	Chemistry, Inorganic \& Nuclear                  &  216,507  & 16,162 & 6,648  \\ \hline
	Obstetrics \& Gynecology                         &  224,383  & 4,221  & 1,634  \\ \hline
	Electrochemistry                                 &  202,755  & 15,001 & 6,253  \\ \hline
	Urology \& Nephrology                            &  217,652  & 3,986  & 1,470  \\ \hline
	Nutrition \& Dietetics                           &  210,908  & 8,264  & 3,596  \\ \hline
	Physics, Particles \& Fields                     &  209,743  & 8,488  & 3,525  \\ \hline
	Engineering, Biomedical                          &  205,175  & 11,495 & 4,725  \\ \hline
	Mechanics                                        &  190,388  & 14,404 & 6,351  \\ \hline
	Engineering, Environmental                       &  190,089  & 13,884 & 6,356  \\ \hline
	Metallurgy \& Metallurgical Engineering          &  192,307  & 12,245 & 5,296  \\ \hline
	Spectroscopy                                     &  192,772  & 11,384 & 4,676  \\ \hline
	Respiratory System                               &  201,044  & 4,160  & 1,484  \\ \hline
	Virology                                         &  196,141  & 6,103  & 2,410  \\ \hline
	Pathology                                        &  194,008  & 4,604  & 1,689  \\ \hline
	Engineering, Mechanical                          &  184,360  & 10,114 & 4,295  \\ \hline
	Veterinary Sciences                              &  188,316  & 6,598  & 2,710  \\ \hline
	Meteorology \& Atmospheric Sciences              &  170,201  & 16,887 & 7,570  \\ \hline
	Nuclear Science \& Technology                    &  183,571  & 7,730  & 3,157  \\ \hline
	Mathematics, Applied                             &  174,821  & 10,513 & 4,790  \\ \hline
	Computer Science, Interdisciplinary Applications &  171,857  & 11,701 & 5,684  \\ \hline
	Marine \& Freshwater Biology                     &  162,891  & 13,465 & 5,954  \\ \hline
	Water Resources                                  &  164,469  & 11,581 & 5,253  \\ \hline
	Orthopedics                                      &  163,871  & 2,982  & 1,191  \\ \hline
	Engineering, Civil                               &  154,356  & 9,313  & 4,348  \\ \hline
	Crystallography                                  &  148,793  & 10,613 & 4,363  \\ \hline
	Geochemistry \& Geophysics                       &  141,171  & 12,737 & 5,482  \\ \hline
	Transplantation                                  &  149,401  & 3,088  & 1,062  \\ \hline
	Health Care Sciences \& Services                 &  143,766  & 3,714  & 1,616  \\ \hline
	Materials Science, Coatings \& Films             &  134,968  & 9,339  & 3,845  \\ \hline
	Physics, Fluids \& Plasmas                       &  126,520  & 10,891 & 4,735  \\ \hline
	Computer Science, Information Systems            &  133,149  & 5,639  & 2,669  \\ \hline
	Dentistry, Oral Surgery \& Medicine              &  131,624  & 4,108  & 1,345  \\ \hline
	Ophthalmology                                    &  132,767  & 2,897  & 1,135  \\ \hline
	Telecommunications                               &  127,165  & 5,334  & 2,274  \\ \hline
	Dermatology                                      &  130,731  & 2,317  &  897   \\ \hline
	Zoology                                          &  122,289  & 8,191  & 3,349  \\ \hline
	Agronomy                                         &  123,303  & 6,429  & 2,904  \\ \hline
	Materials Science, Biomaterials                  &  120,169  & 7,821  & 3,099  \\ \hline
	Computer Science, Artificial Intelligence        &  118,095  & 6,937  & 3,454  \\ \hline
	Sport Sciences                                   &  119,920  & 4,270  & 1,791  \\ \hline
	Behavioral Sciences                              &  115,975  & 6,884  & 2,975  \\ \hline
	Parasitiology                                    &  117,331  & 5,289  & 2,111  \\ \hline
	Physics, Mathematical                            &  109,077  & 10,658 & 4,918  \\ \hline
	Critical Care Medicine                           &  120,190  & 2,440  &  928   \\ \hline
	Economics                                        &  115,749  & 4,467  & 2,140  \\ \hline
	Mathematics                                      &  113,815  & 4,982  & 2,231  \\ \hline
	Physics, Nuclear                                 &  107,096  & 5,943  & 2,564  \\ \hline
	Rheumatology                                     &  111,076  & 2,555  & 1,018  \\ \hline
	Engineering, Multidisciplinary                   &  104,843  & 5,960  & 2,741  \\ \hline
	Psychology                                       &  105,450  & 5,376  & 2,348  \\ \hline
	Evolutionary Biology                             &  98,061   & 10,339 & 4,580  \\ \hline
	Psychology, Clinical                             &  105,390  & 3,641  & 1,528  \\ \hline
	Rehabilitation                                   &  105,498  & 3,078  & 1,269  \\ \hline
	Developmental Biology                            &  103,322  & 4,509  & 1,719  \\ \hline
	Geriatrics \& Gerontology                        &  103,823  & 3,751  & 1,520  \\ \hline
	Reproductive Biology                             &  103,928  & 3,514  & 1,403  \\ \hline
	Thermodynamics                                   &  98,378   & 5,671  & 2,557  \\ \hline
	Mathematical \& Computational Biology            &  96,268   & 6,745  & 3,242  \\ \hline
	Oceanography                                     &  91,393   & 8,484  & 3,976  \\ \hline
	Mathematics, Interdisciplinary Applications      &  93,318   & 6,649  & 3,083  \\ \hline
	Agriculture, Multidisciplinary                   &  90,952   & 5,362  & 2,534  \\ \hline
	Computer Science, Software Engineering           &  91,754   & 4,207  & 1,985  \\ \hline
	Agriculture, Dairy \& Animal Science             &  91,979   & 3,448  & 1,565  \\ \hline
	Operations Research \& Management Science        &  87,893   & 5,125  & 2,440  \\ \hline
	Otorhinolaryngology                              &  90,801   & 1,428  &  457   \\ \hline
	Psychology, Multidisciplinary                    &  86,782   & 3,081  & 1,433  \\ \hline
	Psychology, Experimental                         &  80,966   & 4,868  & 2,179  \\ \hline
	Computer Science, Theory \& Methods              &  81,640   & 3,980  & 1,928  \\ \hline
	Anesthesiology                                   &  83,430   & 1,482  &  561   \\ \hline
	Automation \& Control Systems                    &  77,679   & 4,673  & 2,197  \\ \hline
	Geography, Physical                              &  73,991   & 6,662  & 3,069  \\ \hline
	Fisheries                                        &  75,959   & 4,590  & 2,014  \\ \hline
	Materials Science, Ceramics                      &  73,053   & 4,825  & 2,083  \\ \hline
	Nursing                                          &  77,911   & 1,378  &  666   \\ \hline
	Engineering, Manufacturing                       &  72,987   & 3,530  & 1,588  \\ \hline
	Medical Laboratory Technology                    &  74,890   & 1,628  &  604   \\ \hline
	Health Policy \& Services                        &  74,082   & 1,821  &  828   \\ \hline
	Statistics \& Probability                        &  70,142   & 3,527  & 1,598  \\ \hline
	Management                                       &  70,097   & 3,070  & 1,461  \\ \hline
	Soil Science                                     &  66,746   & 5,322  & 2,320  \\ \hline
	Entomology                                       &  65,502   & 3,811  & 1,537  \\ \hline
	Psychology, Developmental                        &  66,871   & 2,163  &  857   \\ \hline
	Tropical Medicine                                &  66,407   & 1,977  &  800   \\ \hline
	Acoustics                                        &  64,185   & 3,204  & 1,403  \\ \hline
	Neuroimaging                                     &  61,651   & 3,452  & 1,533  \\ \hline
	Education \& Educational Research                &  63,173   & 1,557  &  711   \\ \hline
	Emergency Medicine                               &  64,179   &  823   &  349   \\ \hline
	Biodiversity Conservation                        &  56,081   & 5,941  & 2,614  \\ \hline
	Forestry                                         &  56,561   & 4,666  & 2,081  \\ \hline
	Integrative \& Complementary Medicine            &  61,011   & 1,402  &  614   \\ \hline
	Computer Science, Hardware \& Architecture       &  59,114   & 2,255  &  980   \\ \hline
	Environmental Studies                            &  57,358   & 3,282  & 1,683  \\ \hline
	CELL \& TISSUE ENGINEERING                       &  58,880   & 2,407  &  906   \\ \hline
	Construction \& Building Technology              &  56,236   & 2,872  & 1,377  \\ \hline
	Agricultural Engineering                         &  54,673   & 3,241  & 1,472  \\ \hline
	Engineering, Industrial                          &  54,623   & 2,822  & 1,319  \\ \hline
	Substance Abuse                                  &  53,831   & 1,544  &  709   \\ \hline
	Allergy                                          &  53,964   & 1,331  &  486   \\ \hline
	Gerontology                                      &  49,343   & 1,612  &  653   \\ \hline
	Remote Sensing                                   &  45,947   & 3,537  & 1,744  \\ \hline
	Medical Informatics                              &  48,041   & 1,797  &  783   \\ \hline
	Business                                         &  45,039   & 1,731  &  766   \\ \hline
	Materials Science, Composites                    &  42,725   & 2,807  & 1,233  \\ \hline
	Horticulture                                     &  43,288   & 2,049  &  927   \\ \hline
	Imaging Science \& Photographic Technology       &  40,866   & 3,022  & 1,429  \\ \hline
	Psychology, Social                               &  42,662   & 1,673  &  691   \\ \hline
	Mineralogy                                       &  39,911   & 3,224  & 1,327  \\ \hline
	Education, Scientific Disciplines                &  41,820   & 1,181  &  594   \\ \hline
	Anatomy \& Morphology                            &  39,503   & 1,748  &  648   \\ \hline
	Social Sciences, Interdisciplinary               &  39,062   & 1,293  &  625   \\ \hline
	Social Sciences, Biomedical                      &  39,252   & 1,078  &  465   \\ \hline
	Psychology, Applied                              &  37,772   & 1,381  &  551   \\ \hline
	Limnology                                        &  33,200   & 3,217  & 1,399  \\ \hline
	Mining \& Mineral Processing                     &  34,398   & 1,824  &  787   \\ \hline
	Mycology                                         &  33,552   & 1,825  &  764   \\ \hline
	Geology                                          &  31,450   & 2,500  & 1,004  \\ \hline
	Medicine, Legal                                  &  32,472   &  869   &  400   \\ \hline
	Information Science \& Library Science           &  31,160   & 1,663  &  894   \\ \hline
	Materials Science, Characterization, Testing     &  30,062   & 1,767  &  808   \\ \hline
	Paleontology                                     &  28,681   & 2,205  &  934   \\ \hline
	Engineering, Aerospace                           &  30,096   & 1,189  &  483   \\ \hline
	Transportation Science \& Technology             &  29,061   & 1,476  &  645   \\ \hline
	Engineering, Geological                          &  28,542   & 1,693  &  772   \\ \hline
	Business, Finance                                &  29,634   &  883   &  333   \\ \hline
	Geography                                        &  26,731   & 1,597  &  763   \\ \hline
	Linguistics                                      &  26,786   &  853   &  370   \\ \hline
	Microscopy                                       &  24,798   & 1,979  &  838   \\ \hline
	Sociology                                        &  25,846   &  913   &  397   \\ \hline
	Anthropology                                     &  24,749   & 1,263  &  626   \\ \hline
	Family Studies                                   &  25,566   &  519   &  218   \\ \hline
	Psychology, Biological                           &  23,561   & 1,399  &  660   \\ \hline
	AUDIOLOGY \& SPEECH-LANGUAGE PATHOLOGY           &  23,155   & 1,025  &  418   \\ \hline
	Psychology, Educational                          &  23,105   &  639   &  277   \\ \hline
	Materials Science, Textiles                      &  21,540   & 1,017  &  470   \\ \hline
	Robotics                                         &  21,017   & 1,111  &  509   \\ \hline
	Political Science                                &  20,684   &  650   &  273   \\ \hline
	Planning \& Development                          &  20,237   &  819   &  407   \\ \hline
	Social Work                                      &  20,765   &  403   &  180   \\ \hline
	Transportation                                   &  19,266   & 1,032  &  497   \\ \hline
	Social Sciences, Mathematical Methods            &  18,818   &  858   &  391   \\ \hline
	Ergonomics                                       &  18,782   &  837   &  401   \\ \hline
	Communication                                    &  18,702   &  543   &  234   \\ \hline
	Education, Special                               &  18,090   &  459   &  221   \\ \hline
	Archaeology                                      &  17,347   &  849   &  417   \\ \hline
	PRIMARY HEALTH CARE                              &  17,766   &  502   &  251   \\ \hline
	Criminology \& Penology                          &  16,556   &  319   &  136   \\ \hline
	Urban Studies                                    &  15,062   &  710   &  332   \\ \hline
	Computer Science, Cybernetics                    &  14,705   &  768   &  382   \\ \hline
	Engineering, Petroleum                           &  14,808   &  395   &  177   \\ \hline
	Women’s Studies                                  &  14,437   &  253   &  102   \\ \hline
	Engineering, Ocean                               &  13,058   &  871   &  393   \\ \hline
	Hospitality, Leisure, Sport \& Tourism           &  13,143   &  494   &  256   \\ \hline
	International Relations                          &  12,663   &  385   &  171   \\ \hline
	Ornithology                                      &  11,503   & 1,011  &  427   \\ \hline
	Ethics                                           &  12,496   &  309   &  127   \\ \hline
	Andrology                                        &  11,678   &  289   &  113   \\ \hline
	Social Issues                                    &  11,581   &  310   &  126   \\ \hline
	Language \& Linguistics Theory                   &  11,469   &  252   &  116   \\ \hline
	Materials Science, Paper \& Wood                 &  10,562   &  627   &  279   \\ \hline
	Public Administration                            &  10,454   &  355   &  168   \\ \hline
	Law                                              &  10,358   &  176   &   73   \\ \hline
	Psychology, Mathematical                         &   7,344   &  401   &  182   \\ \hline
	History \& Philosophy of Science                 &   6,958   &  206   &   98   \\ \hline
	Medical Ethics                                   &   7,016   &  174   &   71   \\ \hline
	Demography                                       &   6,361   &  167   &   77   \\ \hline
	Engineering, Marine                              &   6,202   &  259   &  127   \\ \hline
	Agricultural Economics \& Policy                 &   5,954   &  267   &  143   \\ \hline
	Industrial Relations \& Labor                    &   5,739   &  152   &   72   \\ \hline
	History                                          &   4,887   &   69   &   41   \\ \hline
	Area Studies                                     &   4,222   &   84   &   40   \\ \hline
	Religion                                         &   4,239   &   58   &   25   \\ \hline
	LOGIC                                            &   3,715   &  209   &  117   \\ \hline
	Ethnic Studies                                   &   3,918   &   80   &   29   \\ \hline
	Philosophy                                       &   3,715   &   64   &   39   \\ \hline
	Art                                              &   3,190   &  162   &   84   \\ \hline
	History of Social Sciences                       &   3,086   &   92   &   52   \\ \hline
	Humanities, Multidisciplinary                    &   3,043   &   52   &   31   \\ \hline
	Psychology, Psychoanalysis                       &   2,501   &   52   &   21   \\ \hline
	CULTURAL STUDIES                                 &   2,361   &   34   &   16   \\ \hline
	Music                                            &   1,998   &   47   &   20   \\ \hline
	Architecture                                     &   1,843   &   77   &   48   \\ \hline
	Literature                                       &   1,567   &   22   &   11   \\ \hline
	Film, Radio, Television                          &    795    &   20   &   8    \\ \hline
	Asian Studies                                    &    704    &   5    &   1    \\ \hline
	Theater                                          &    506    &   2    &   0    \\ \hline
	Medieval \& Renaissance Studies                  &    253    &   3    &   1    \\ \hline
	Literary Reviews                                 &    160    &   0    &   0    \\ \hline
	Dance                                            &    128    &   0    &   0    \\ \hline
	Literature, Romance                              &    122    &   1    &   1    \\ \hline
	Literary Theory \& Criticism                     &    91     &   0    &   0    \\ \hline
	Folklore                                         &    87     &   2    &   1    \\ \hline
	Classics                                         &    85     &   0    &   0    \\ \hline
	Literature, German, Dutch, Scandinavian          &    67     &   0    &   0    \\ \hline
	Literature, British Isles                        &    62     &   0    &   0    \\ \hline
	Literature, American                             &    53     &   0    &   0    \\ \hline
	Literature, African, Australian, Canadian        &    27     &   0    &   0    \\ \hline
	Poetry                                           &     5     &   0    &   0    \\ \hline
\end{longtable}